\documentclass[reprint, prd,floatfix,longbibliography]{revtex4-2}
\usepackage{graphicx}% Include figure files
\usepackage{dcolumn}% Align table columns on decimal point
\usepackage{comment}
\usepackage{bm}% bold math
\usepackage{hyperref}
\usepackage{enumitem}
\usepackage{xcolor}
\usepackage{physics}
\usepackage{amsmath}
\usepackage{nicefrac}
\usepackage{lineno, blindtext}
\newcommand{\cor}{\textcolor{black}}

\begin{document}
    % \begin{linenumbers}
    
    \title{Emergent Chirality and Current Generation}
    \author{Satyanu Bhadra}
    \affiliation{Department of Condensed Matter Physics and Materials Science, \\ Tata Institute of Fundamental Research, Mumbai 400005, India}
    \author{Shankar Ghosh}%
    \affiliation{Department of Condensed Matter Physics and Materials Science, \\ Tata Institute of Fundamental Research, Mumbai 400005, India}%
    \author{Shamik Gupta}
    \affiliation{Department of Physics, Ramakrishna Mission Vivekananda
        Educational and Research Institute, Belur Math, Howrah 711202, India,\\ Regular Associate, Quantitative Life Sciences Section, ICTP - The Abdus Salam International Centre for Theoretical Physics, Strada Costiera 11, 34151 Trieste, Italy}
    \date{\today}% It is always \today, today,
    % but any date may be explicitly specified
    
    \begin{abstract}
        We investigate the phenomenon of producing vibration-induced rotational motion in a cylinder filled with achiral rods and being vibrated by an external drive. The arrangement of the rods develops chirality as a result of the interaction of gravity and steric hindrance, which is responsible for the induced motion. A two-rod arrangement is sufficient to generate a persistent motion. The average angular velocity $\langle \omega \rangle$ of the rods at long times varies non-monotonically with the packing fraction $\phi$ for a given drive strength $\Gamma$. Though the precise nature of the variation of $\langle \omega \rangle$ with $\phi$ depends on the details of the interaction between individual rods, its general characteristics hold true for rods of various materials and geometries. A stochastic model based on the asymmetric simple exclusion process helps in understanding the key features of our experiments.
    \end{abstract}
    
    \pacs{Valid PACS appear here}% PACS, the Physics and Astronomy
    \maketitle
    
    It is common knowledge that when a system is subjected to an external field that causes a current to flow in it, it behaves differently from when it is not. Presence of current in the system implies that the underlying dynamics no longer satisfies time-reversal symmetry. The external field induces biased transitions between configurations by effectively rendering a subset of configurations in the energy landscape to be higher in energy than the others. The scenario described above is best visualized in a system of interacting particles, which in the thermodynamic limit develops an effective single-particle potential energy landscape that is tilted due to the application of an external field to have two minima separated by an energy barrier. An alternative method of generating current would be to use extended objects that break structural symmetry, so that when a field is applied, these objects move in a preferred direction. Chiral objects are an example of the aforementioned type. These objects convert linear momentum to angular momentum and thus twist and turn as they attempt to move under the influence of an external field \cite{notchedstick, altshuler2013vibrot, scholz2016ratcheting, Scholzeabf8998, case2014rattleback, NANDArattleback, kondorattle, tsai2005chiral, CICCONOFRI2015233, giomi2013swarming, scholz2018rotating, Vibrot1, V_lkel_2020, liu2020oscillating}.
    A question of pertinent interest is then: Can an external field drive current in a system of achiral particles interacting only through steric hindrance, and that too not in the thermodynamic limit but for a system of a few particles?
    
\begin{figure}[t]
        \includegraphics[width=.95\linewidth]{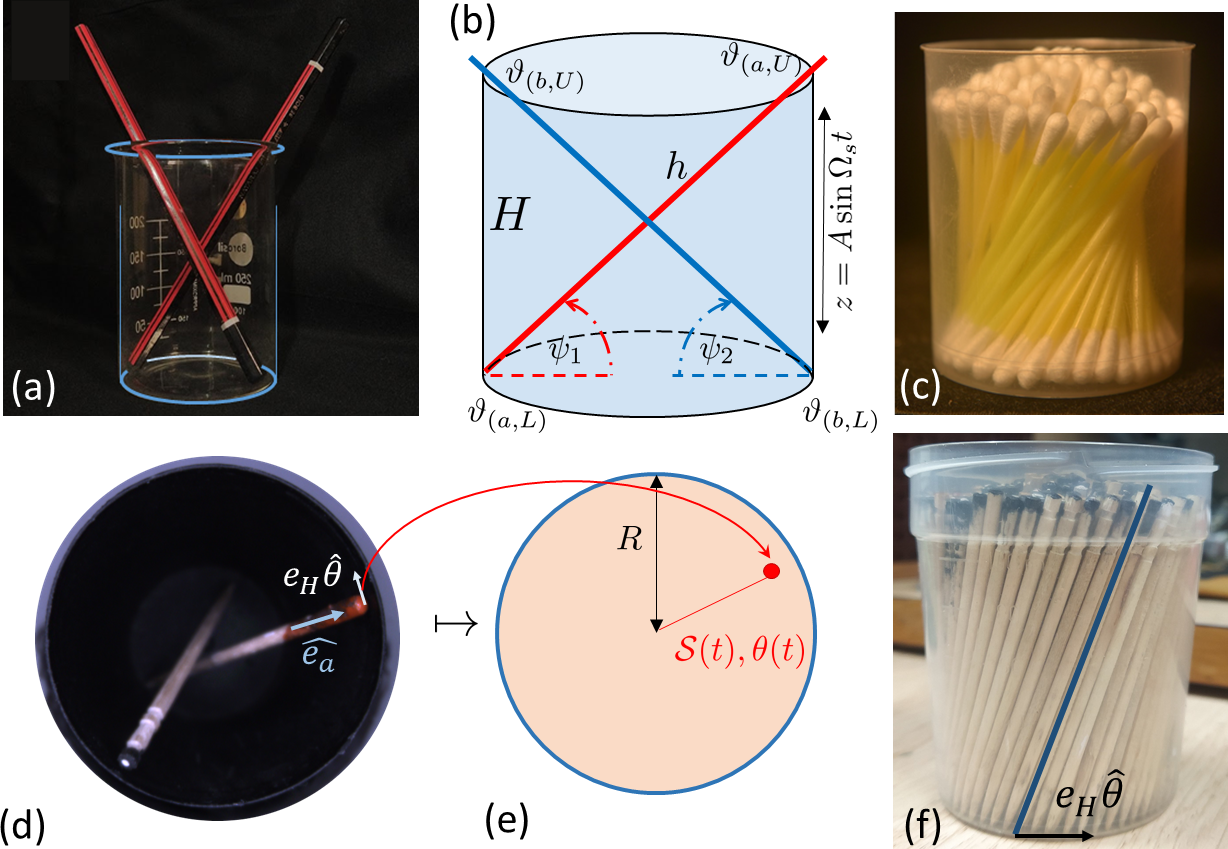}
        \caption{ (a) A typical experimental scenario in which two pencils are placed in a container. These pencils can be made to move in one direction by tapping the container. The relative arrangement of the pencils determines the direction. (b) A schematic representation of the experiment, including all relevant coordinates. (c) A cylinder of earbuds generates a doubly-ruled hyperbolic surface. Take note of the profile of the earbuds' top-ends; the rods on the inside are straighter than the ones on the outside.
            (d) A typical snapshot of the system of rods being driven. From the top view, we can track the top of the tagged particle, colored red. The unit vector $\hat{e}_a$ lies along the rod, with the tangential vector being $e_H \hat{\theta}$. (e) A schematic showing the mapping of the $x-y$ coordinates at time $t$ of the tagged particle top in Fig 1(d) onto a disk geometry, with relevant coordinates $\mathcal{S}(t)$ and $\theta(t)$ (f) The component $e_H$ can be measured by the angle made by a rod with the wall of the container. A single toothpick and corresponding $e_H$ is highlighted.}
        \label{fig:expt}
    \end{figure}

    The aforesaid question is answered in the affirmative in this work: In a system of rods confined to a cylindrical container, we show that vertical shaking of the cylinder can spontaneously generate a chirality in the system, leading to a rotational current whose direction is determined by the handedness of the developed chirality. The experimental setup is depicted in Fig.~\ref{fig:expt}. The radius of the cylinder is $R$, while its height is $H$. This cylinder is filled with $N$ identical rods of length $h$ and width $2r$ that are achiral. To impart regulated vibrations to the rods, we employ an electromagnetic shaker that vibrates the cylinder in the vertical (i.e., in the $z$) direction with frequency $\Omega_s$ and amplitude $A$. The parameter $\Gamma \equiv A \Omega_{s}^2/g$ sets the strength of this external drive. We find that a rotational current is generated in the system so long as we have $r<R$, $h>2R$, and the parameter $\Gamma$ is less than a critical value. We performed experiments with objects of varying $h$ and found that experimental results varied weakly. If $h >> H$, the object can topple over the container when shaken. When subject to the drive, and when one has a few rods in the cylinder, the system settles into a chiral arrangement with the ends of the rods resting on the top and the bottom rim of the cylindrical container. Despite the fact that the Euclidean geometry of the rods is non-chiral, presence of gravity lowers the symmetry group associated with the geometry of the rods. Chirality arises from the fact that the reduced group has fewer connected components.
    
    We now discuss the coordinate description of the problem, for the representative two-rod case where the rods are in contact. We label the rod on top at the point of contact as $a$ and the one below as $b$.
    In terms of cylindrical polar coordinates, the end points of rod $a$ are $[(R,\vartheta_{(a,L)}, 0),(R,\vartheta_{(a,U)}, H)]$ and the end points of rod $b$ are $[(R,\vartheta_{(b,L)}, 0),(R,\vartheta_{(b,U)}, H)]$. Here, the $L$ and $U$ indices in the subscript of $\vartheta$ describe the lower and upper end points of the rods. The lower  end of the rods rests on the bottom rim of the container, while the upper end of the rods rests on the upper rim. If rod $a$ needs to be turned in a clockwise (respectively, anticlockwise) direction so as to make it parallel to rod $b$, we identify the structure with negative (respectively, positive) chirality. Zero chirality represents configurations where the rods are parallel to each other.
    
    In the following, we discuss in turn our experimental observations made with one, two, and more than two rods in our setup. To study the dynamics of the rods in time, we track the motion of a single tagged rod. From the top view, the motion of the top of this tagged rod appears to correspond to the movement of a point on a disk whose center lies along the axis of the container. It is thus natural to characterize the motion of this rod in terms of polar coordinates $\mathcal{S}(t) , \theta(t)$. The angular coordinate $\theta(t)$ describes the instantaneous angular position of this tagged rod on this disk (see Fig.~\ref{fig:expt}, panels (d),(e)). From the time series $\theta(t)$, we construct the cumulative angular distance $\Theta(t)$, where $\theta(t)= {\rm mod}|\Theta(t),2\pi|$. The tangential component $e_H$ of the axial vector in the direction of $\hat{\theta}$ changes as the container fills up with more rods, and is correlated with the angular velocity of the rods; This is discussed in detail later in the paper. The tangential component can be easily computed from images such as Fig.~\ref{fig:expt}, panels (d) and (f).
    
    \begin{figure}[t]
        \includegraphics[width=.95\linewidth]{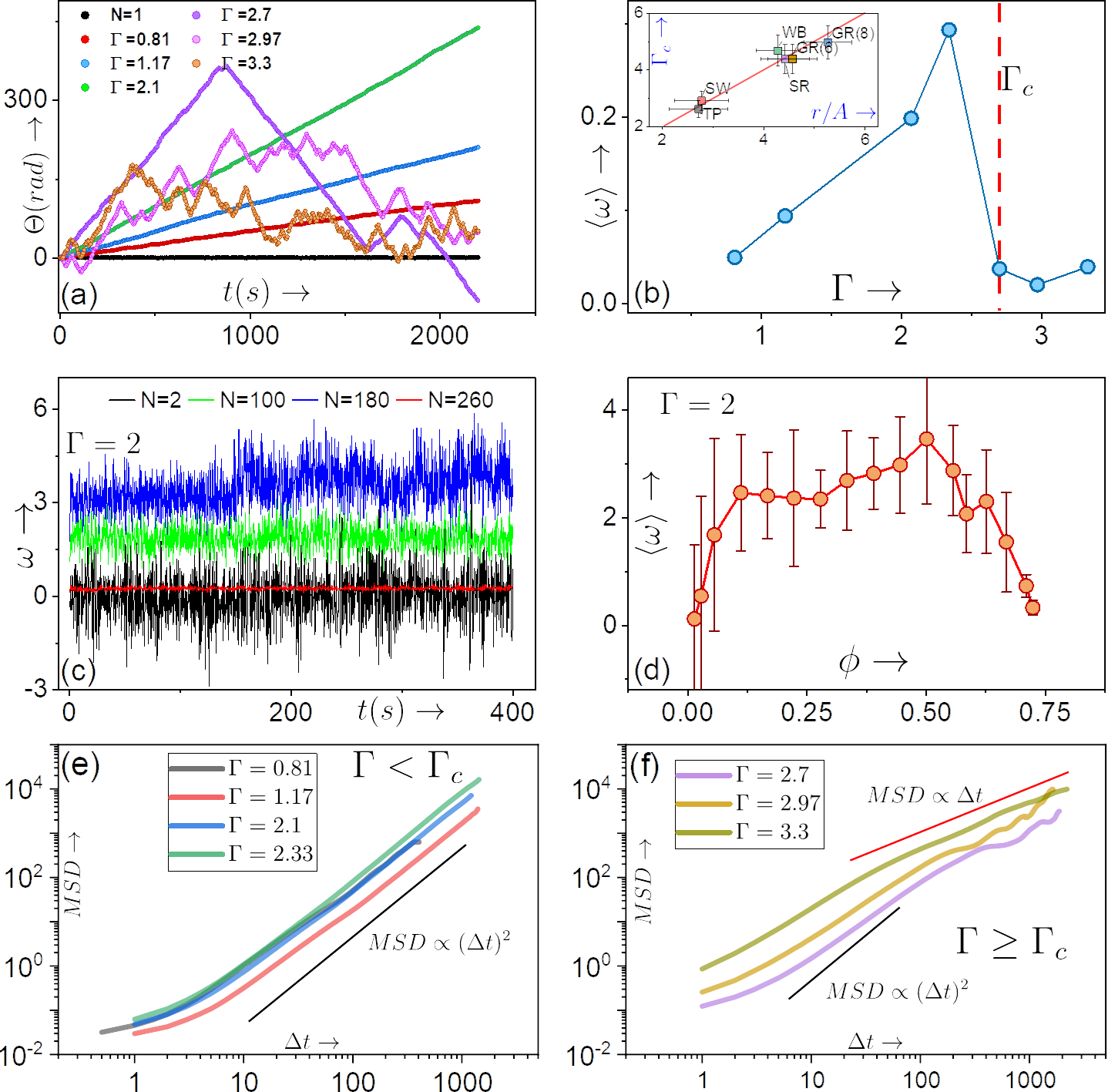}
        
        \caption{(a) Time trace of the angular distance $\Theta(t)$ of a tagged rod in a system of $N$ rods, for $N=2$ and various values of $\Gamma$. The case $N=1$ is also shown.
            (b) Long-time average of the angular velocity, denoted by $\langle\omega\rangle$, increases with $\Gamma$ until $\Gamma_{c}$, after which continuous chirality toggling causes $\langle\omega\rangle$ to plummet.
            Inset: The variation of $\Gamma_c$ vs $r/A$ for different materials of varying thickness. The red line represents $\Gamma_c = r/A$.
            (c) The toothpicks move noisily, with the amount of noise decreasing as $N$ increases. This is seen in the time trace of the angular velocity $\omega$.
            (d) The variation of $\langle \omega \rangle$ with packing fraction $\phi$ for toothpicks ($r=2.2$mm, $h=60$mm, $R=45$mm, $H=75$ mm). Note that the error bars reduce with increase in $\phi$.
            The panels (e) and (f) show the variation of $MSD$ vs lag time $\Delta t$ for $\Gamma < \Gamma_c$ and $\Gamma \ge \Gamma_c$ respectively. }
        \label{fig:2particle}
    \end{figure}
    
    \paragraph{ Observations for $N=1$:}
    Chirality is an emergent property that arises from the arrangement of two or more rods. The rods used here are not by themselves chiral. On being driven, a single rod does not exhibit any persistent angular motion (see Fig.~\ref{fig:2particle}(a), $N=1$ ).

    \paragraph{ Observations for $N=2$:}
    In terms of $\Theta(t)$, the trajectory of a tagged rod for various values of $\Gamma $ is shown in Fig.~\ref{fig:2particle}(a). Occurrence of a persistent rotational current necessitates the use of at least two rods. The direction of rotation of two rods is determined by the instantaneous chirality in the system. Constant toggling between states with two different signs of chirality results in a continuous current reversal and a concomitant change in sign of the slope of $\Theta(t)$ versus $t$. The rate of toggling increases as $\Gamma$ increases, and there exists a $\Gamma_{c}$ beyond which the long-time average $\langle \omega \rangle$ of the angular velocity starts to drop, as seen in Fig.~\ref{fig:2particle}(b).  By measuring the mean square displacement $MSD$ as a function of the lag time $\Delta t$, we see that for $\Gamma < \Gamma_c$, the motion is completely ballistic ($MSD \propto (\Delta t)^2$) as seen in Fig \ref{fig:2particle}(e). For $\Gamma \ge \Gamma_c$, the motion is ballistic at shorter timescales, and makes a crossover into diffusive motion ($MSD \propto \Delta t$) for longer timescales, as seen in \ref{fig:2particle}(f).
    
    \begin{figure}[t]
        \includegraphics[width=.95\linewidth]{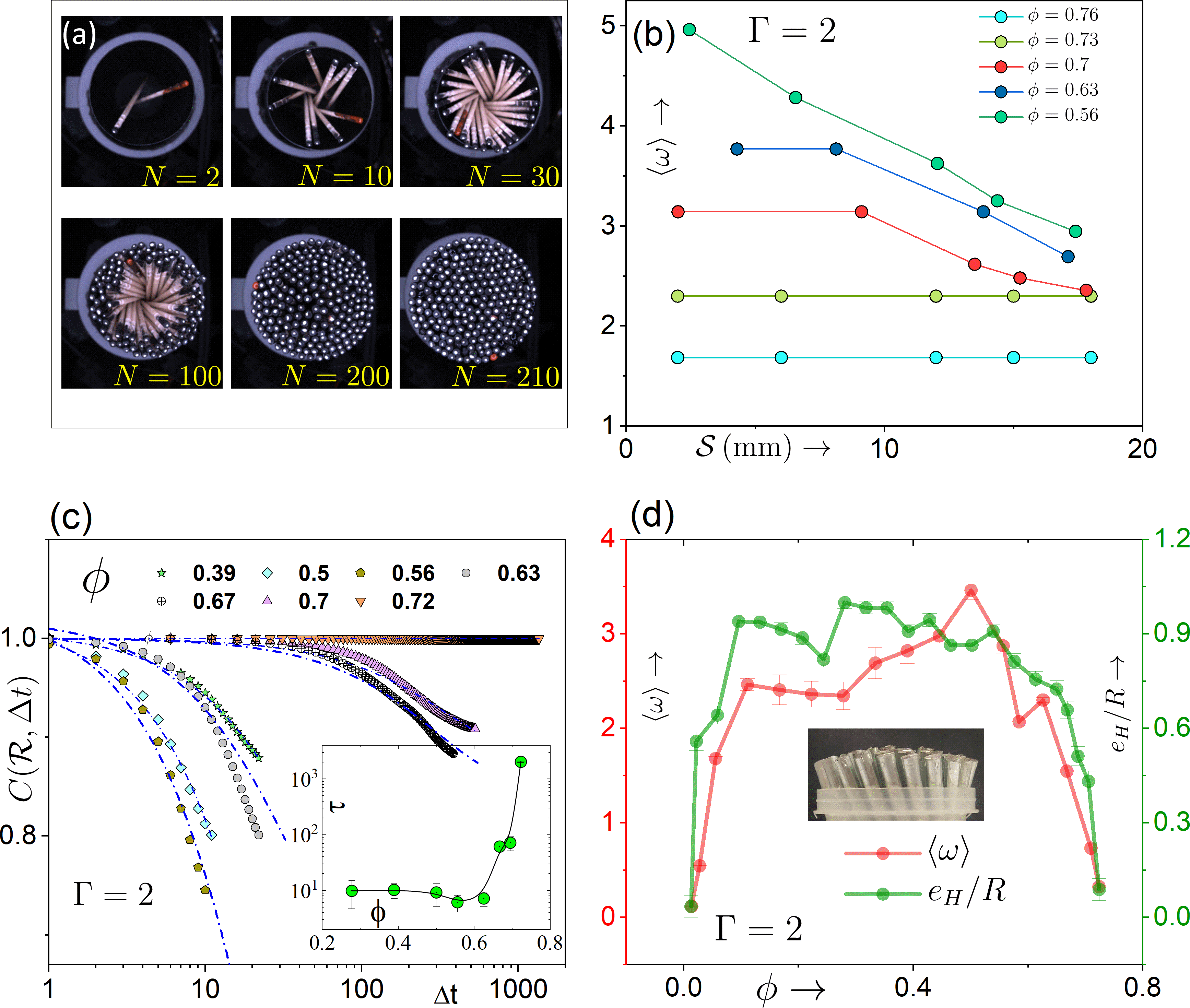}
        \caption{(a) Snapshots of rod packing with increase in number $N$ of rods. As we load more rods, the tendency to form clusters rather than equidistant structures increases. Well-defined shells of rods are observed beyond a certain number of rods. (b) Variation of long-time average $\langle \omega \rangle$ of the angular velocity with radial distance $\mathcal{S}$, measured along the radius of the container, for various packing fractions $\phi$, highlighting the presence of a velocity gradient as one travels from the container's center to the edge. This gradient decreases with increasing $\phi$. (c) The variation of the autocorrelation function $C(\mathcal{R}, \Delta t)$ for various $\phi$ and respective exponential fits $g(t)=\exp(-t/\tau)$. Inset: Variation of $\tau$ with $\phi$. (d) Variation of $e_{H}/R$ and $\langle \omega \rangle$ with $\phi$. All experiments were performed using toothpicks ($r=2.2$mm, $h=60$mm, $H=75$mm, $R=37$mm).}
        \label{fig:ParticleN}
    \end{figure}
    
    \paragraph{Observations for $N>2$:}
    As the number of rods in the container increases beyond two, the arrangement of the rods evolves from equidistant regular structures (see the case $N=2$ in Fig.~\ref{fig:ParticleN}(a)) to a single shell that resembles a doubly-ruled hyperbolic surface (see the case $N=30$ in Fig.~\ref{fig:ParticleN}(a) and Fig. \ref{fig:expt}(c)). When the number is further increased, multiple concentric shells are formed. For low values of the packing fraction $\phi \equiv N (r/R)^2$, the rods move from one shell to the next as these shells shear against each other.
    The presence of shearing between shells is demonstrated by a gradient in $\langle \omega \rangle$ as a function of the radial distance $\mathcal{S}$ measured along the radius of the container (Fig. \ref{fig:ParticleN}(b)). As the packing fraction $\phi$ increases, $\langle \omega \rangle$ becomes increasingly independent of $\mathcal{S}$. The motion of the rods appears to become more coherent with increase in $\phi$; $\phi_{\rm rigid}$ refers to the value of $\phi$ at which the shearing between layers completely disappears.
    In order to quantify the onset of coherence in the motion, one may consider studying an appropriate autocorrelation function. To this end, we map the top of two tagged rods onto two points on a disc whose coordinates are $(\mathcal{S}_1(t),\theta_1(t))$ and $(\mathcal{S}_2(t),\theta_2(t))$ and measure the Pythagorean distance between the points. Denoting this distance at time $t$ by $\mathcal{R}(t)$, we have
    $  \mathcal{R}^2 (t) = \mathcal{S}_1^2(t)+\mathcal{S}_2^2(t)-2 \mathcal{S}_1 \mathcal{S}_2 \cos\left(\theta_1(t) - \theta_2(t)\right).
    $
    The autocorrelation function is defined as $C(\mathcal{R}, \Delta t) \equiv \langle\mathcal{R}(t)\mathcal{R}(t+\Delta t)\rangle$ for time lag $\Delta t$.
    For various values of $\phi$, the behavior of the correlation function is presented in Fig. \ref{fig:ParticleN}(c). Because the two tagged rods are rotating within the cylinder, the distance between them is an oscillating function. As a result, the correlation function as a function of $\Delta t$ shows a decay with superimposed small oscillations. Fitting the decaying profile to an exponential yields the correlation time $\tau$. Exponential fit $g(t)=\exp(-t/\tau)$ to the different correlation functions are represented by the dashed-dotted lines in Fig. \ref{fig:ParticleN}(c).
    As $\phi \rightarrow \phi_{\rm rigid}$, the system makes a transition from being floppy, where the rods can move freely with respect to one another, to being rigid, where all the rods move coherently, and correspondingly, the timescale $\tau$ becomes large as seen in Fig. \ref{fig:ParticleN}(c) inset. Similar rigidity transitions have been observed in other experimental systems \cite{Origami_Chen8119,THORPE1983355}.
    
    Apart from the linear drift in $\Theta(t)$, there exist oscillations about the linear function, corresponding to the rattling motion of the rods observed in experiments. For low densities, each rod can access a large section of the configuration space. On being driven by the shaker, the injected energy does not fully contribute to the global motion of the rods, but is used by individual rods to explore nearby configurations in its allowed configuration space. This leads to a noisy, rattling motion of the rods at low densities, making the motion of the individual rods to be essentially random on short timescales (Fig. \ref{fig:2particle}(c)). The rattling of the rods decreases as the number of rods in the container increases since the available configuration space for each rod decreases; the system eventually becomes rigid (Fig. \ref{fig:2particle}(c), (d)).
    
    Each rod is tilted at an agle with respect to the vertical along the wall of the container. We quantify this tilt as the as tangential component $e_H$ of the axial vector lying along $\hat{\theta}$, where $\hat{\theta}$ is the normal vector to the radial vector from the centre of the container. $e_H$ of the rods changes as the container fills up with more rods. For the rod $a$ shown in Fig.\ref{fig:expt}(b) $e_{H} = R \sin(\vartheta_{(a,U)} - \vartheta_{(a,L)})$. The quantities $\langle \omega \rangle$ and $e_{H}$ of a tagged rod are depicted as a function of $\phi$ in Fig. \ref{fig:ParticleN}(d). Since the rods travel with different speeds in different layers, we measure the $\langle \omega \rangle$ of a tagged rod placed in the outermost shell for all experiments. Both $e_{H}$ and $\langle \omega \rangle$ have similar non-monotonic trends as a function of packing fraction $\phi$.
    It is tempting to attribute the increase in $\langle \omega \rangle$ to the corresponding increase in $e_H$.
    The rods in the inner shell, on the other hand, are inclined at a lower angle than those at the edge (Fig. \ref{fig:expt}(c)). As one advances radially inwards, $e_H$ drops, whereas investigations show that inner shells move faster (Fig. \ref{fig:ParticleN}(b)). As a result, $e_H$ cannot be the only element influencing particle speed.
    
\begin{figure}[t]
        \includegraphics[width=.95\linewidth]{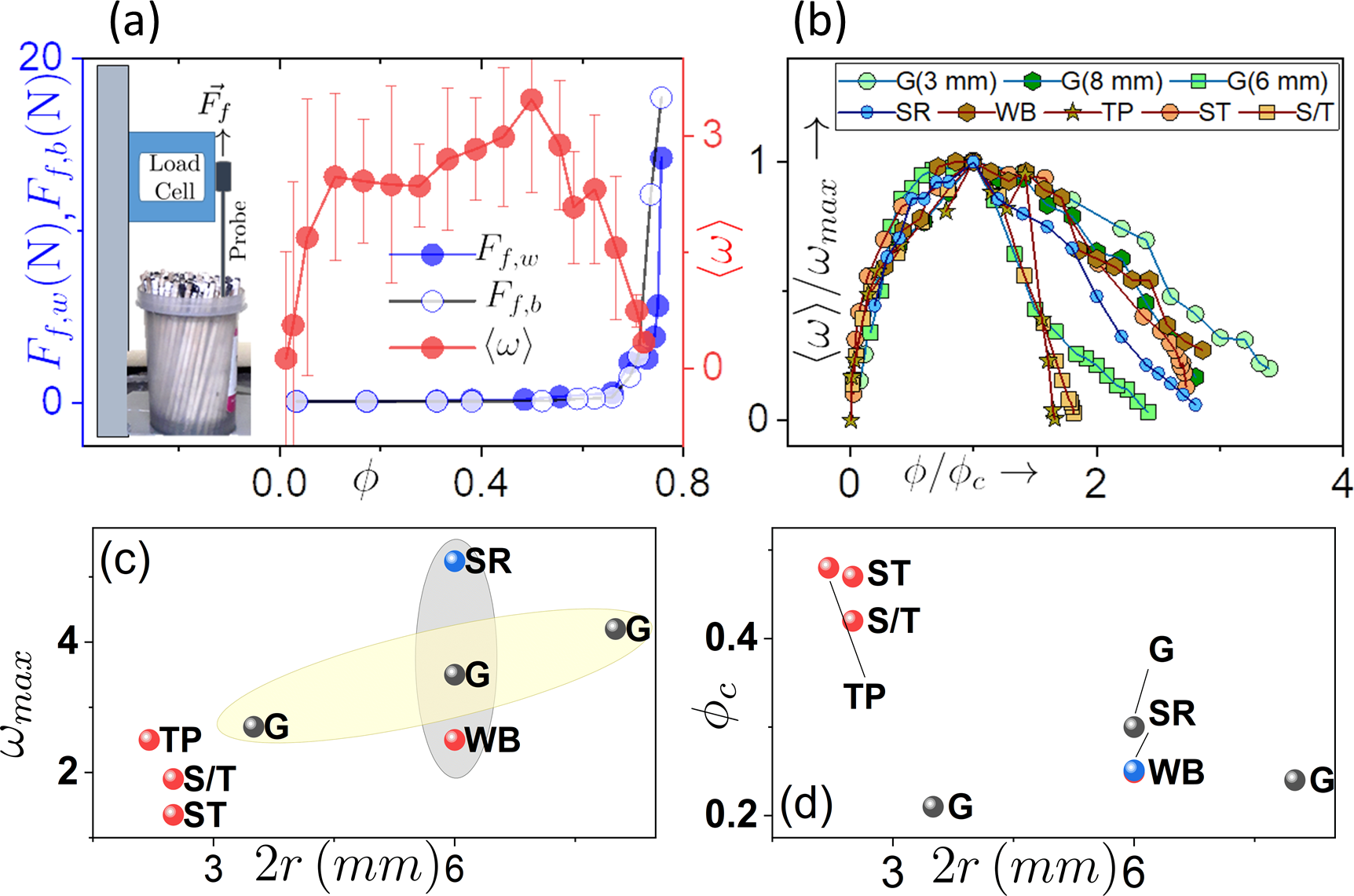}
        
        \caption{(a) Variation of the long-time average $\langle\omega\rangle$ of the angular velocity  ($\Gamma=2$) and the frictional force $F_f$ with packing fraction $\phi$ for toothpicks ($r=2.2$mm, $h=60$mm, $H=75$mm, $R=37$mm). Inset shows a schematic of the setup used to measure $F_f$, where a probe attached to a load cell is inserted into the container full of $N$ rods.
            (b) On plotting $\langle\omega\rangle/\omega_{\rm max}$ vs $\phi/\phi_c$, a good scaling relationship for the rise is observed. (c,d) $\omega_{\rm max}$ is dependent on both the material and the diameter, whereas $\phi_c$ is not affected by the material but by only the diameter.}
        \label{fig:Materials}
    \end{figure}

    The frictional dissipation in the system increases with $\phi$. In the inset of Fig. \ref{fig:Materials}(a), we present a schematic representation of our experimental set-up that measures the force of friction between the rods and the wall of the cylinder. The probe, a toothpick attached to the load cell, is lowered vertically into the container that already has requisite number of toothpicks to make the packing fraction $\phi$. To measure the force of friction between the rods, we insert the probe in the middle of the container, while to measure the frictional force between the rods and the wall, we ensure that the probe continuously brushes against the wall of the container. The process of lowering of the probe is continued till the lower end of the probe comes within 2 mm of the base of the container. The highest frictional force recorded between the probe and the rods is $F_{f,b}$, and the force between the rods and the wall is $F_{f,w}$. Both these measurements are made at the deepest point of insertion, which is 2 mm from the base of the container. The experiment was repeated with different $\phi$ values;
    %for densities
    %$\phi_{\rm jam}$ represents the value of $\phi$ at which the motion of the rods has completely stopped. As $\phi \rightarrow \phi_{\rm jam}$,
    for  $\phi > 0.66$ it becomes increasingly difficult to insert rods into the container, and the value of both $F_{f,b} $ and $F_{f,w}$ rise steeply (Fig. \ref{fig:Materials}(a)). The rise in both $F_{f,b} $ and $F_{f,w} $ and the drop in $\langle \omega \rangle$ occur at values of $\phi$ close to each other. It is to be noted the these two values $\langle \omega \rangle$ and the forces of friction are measured from different experiments. As discussed above, while measuring the force of friction, the system was static. These observations indicate that there are two roles of the frictional force in the system. The inter-particle friction provides constraints and hence is closely associated to the rigidity transition of the system (Fig \ref{fig:ParticleN}(c)). On the other hand the friction between the rod and the wall provides resistance to the particle current. At high packing fraction $\phi$, the friction between wall and the rods exceeds the driving force associated with the current, hence the current goes to zero (Fig \ref{fig:ParticleN}(d)).

    Experiments with different materials of varying sizes and shapes reveal that the nature of the $\langle\omega\rangle$ vs. $\phi$ relationship is always the same. The following materials were used in our setup: Steel rods (SR, $2r$ = 6mm), Toothpicks (TP, $2r$ = 2.2mm), Bamboo skewers with tips (ST, $2r$ = 2.5mm), Bamboo skewers without tips (S/T, $2r$ = 2.5mm), Wood blocks (WB, $2r$ = 6mm), Glass rods (GR, $2r$ = 3.5, 6, 8 mm), Wood blocks (WB, $2r$ = 6mm), Wood blocks (WB, $2r$ = 6mm), Glass rods (GR, $2r$ = 3.5, 6, 8 mm). The ordinate is scaled as $\langle\omega\rangle/\omega_{\rm max}$ and the abscissa is scaled as $\phi/\phi_c$ to compare results from different experiments (Fig. \ref{fig:Materials}(b)). For the region $\phi/\phi_c< 1$, the figure exhibits good scaling collapse of the data. It may be seen from Fig. \ref{fig:Materials}(b) that $\omega_{\rm max}$ is controlled by both the material and the diameter, but $\phi_c$ is only influenced by the diameter and not by the material.
    
    \begin{figure}[t]
        \includegraphics[width=.85\linewidth]{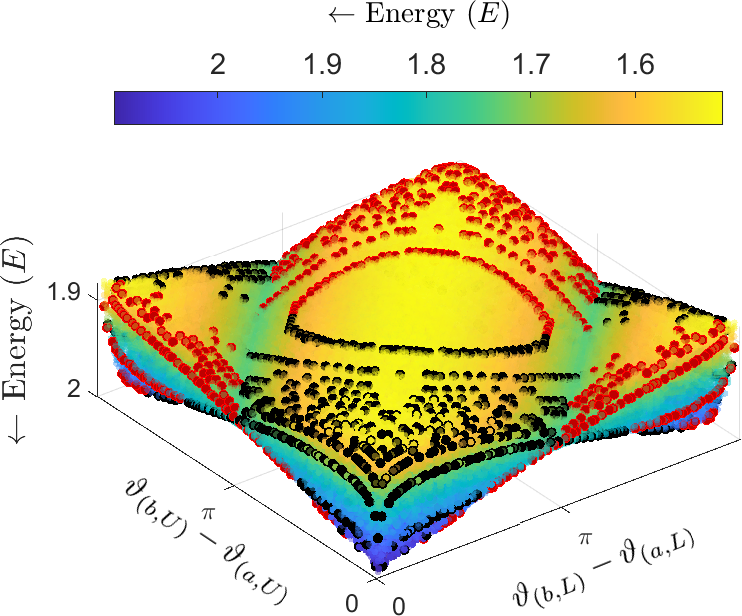}
        \caption{ The graph depicts the energy $E$ of two rods in a cylinder as a function of $\vartheta_{(b,L)}-\vartheta_{(a,L)}$ and $\vartheta_{(b,U)}-\vartheta_{(a,U)}$ for all possible allowed configurations. The red (negative chirality) and black (positive chirality) markers indicate configurations that satisfy the condition $d=2r$, with $d$ and $r$ being the shortest distance at point of contact of two rods, and the radius of the rods, respectively. Refer to Fig. \ref{fig:expt}(b) for the definition of the $\vartheta_i$'s.
        }
        \label{fig:energy}
    \end{figure}
    
    \paragraph{Energetics for the case $N=2$:}
    The energy of each rod is characterized in terms of the tilt angle, and the coordinates of the two ends of a rods are used as its geometrical descriptor.
    The coordinate description of the problem is depicted in detail in Fig. \ref{fig:expt}(b). In these coordinates, the potential energy $E$ of two rods $a$ and $b$ with tilt angles \cor{$\psi_{1}$ and $\psi_{2}$ respectively is
        $(mgh/2) (\sin(\psi_{1})+\sin(\psi_{2}))$}. The sine function of the tilt angle $\psi_{1}$ is $\sin ( \psi_{1})= \left(1 + \left(\frac{2R}{H}\right)^2 \sin^2 \left(\frac{\vartheta_{(a,U)} - \vartheta_{(a,L)}}{2}\right)\right)^{-\frac{1}{2}}$.
    
    In this case, the tilt angles can be expressed in terms of the coordinates of the end points of the rod. When the angular gap between the top and the bottom end of a rod is maximized, the energy for one rod is at its minimum. When a second rod is introduced, the overall energy must be minimized while also considering the steric repulsion of the two rods in contact. This restriction is written as
    $d \ge 2r $, where $d$ is the shortest distance at point of contact between two rods. The scatter plots in Fig.~\ref{fig:energy} illustrate the possible configuration for $R=1$, $H=2$, and $r=0.05$, with the restriction $d \ge 2r$.
    
    The color code represents the total energy $E$ of the two rods. The black and red dots in the figure show the configurations where the rods come into contact with $d=2r$. The black and red dots represent positive and negative chirality, respectively. (i) There are two global minima to be found. Each one has a chirality of its own. As a result, the configuration space corresponding to distinct chiralities is divided by an energy barrier. (ii) Permissible configuration pathways connect these minima. (iii) Because of the energy introduced by the shaker, the rods can explore neighboring configurations.
    
    The configurations of the rods are constrained in the energy diagram for smaller values of $\Gamma$ to a specific chirality. As a result, the rods continue to move in the same direction. The energy supplied by the shaker, on the other hand, is sufficient to toggle the chirality of the arrangement for $\Gamma >\Gamma_c$, and crossing this energy barrier results in current reversal. The quantity $\langle \omega \rangle$ would consequently be zero. The relevant energy barrier $\Delta E$ that divides the chiral states and the energy $2mgA\Gamma_c$ imparted by the shaker may be used to calculate $\Gamma_c$. The motion of one rod ascending over the other describes a simple route in the configuration space that permits the chirality to toggle. This corresponds to the situation where we have $\Delta E= 2mgr$ and $\Gamma_c= \frac{\Delta E}{2Amg}=\frac{r}{A}$. For $A = 0.4$mm, $r= 1.1$mm, we have $\Gamma_c= 2.75$, which is consistent with experimental findings for the data plotted in Fig. \ref{fig:2particle}(b). We have seen that this relation holds for various objects of different materials and diameters as seen in the inset of Fig. \ref{fig:2particle}(b), as long as the height of the object isn't so high that it topples out of the container on being shaken.
    
    \paragraph{Simulations for $N>2$:}
    The geometric model discussed in the foregoing is of use mainly in the two-rod scenario, and its extension to analyze the case of multiple rods is not easy. To proceed, we note that the motion of individual rods on the timescale of our experiments appears stochastic, as mentioned earlier while discussing the rattling motion of rods. This warrants a stochastic dynamical model to explain qualitatively our experimental findings. A few reasonable simplifications are in order, all of which are motivated by experimental observations. When viewed from above, the rods appear as discs moving preferentially in one direction on circular tracks. There are several concentric tracks available for the rods to move in, and when allowed, the rods can move from one track to another. Moreover, manufacturing defects in the cylindrical container lead to the rods speeding up and slowing down at different random regions of the container, leading to a quenched disorder in the system. Steric hindrance forbids two discs to be on top of one another. All these aspects may be accounted for by considering a lattice model of hard-core discs, wherein the separation between neighboring lattice sites is given by the diameter of a disc and hard-core constraint allows only one disc to be placed centered on a lattice site.
    
\begin{figure}[t]
        \includegraphics[width=.95\linewidth]{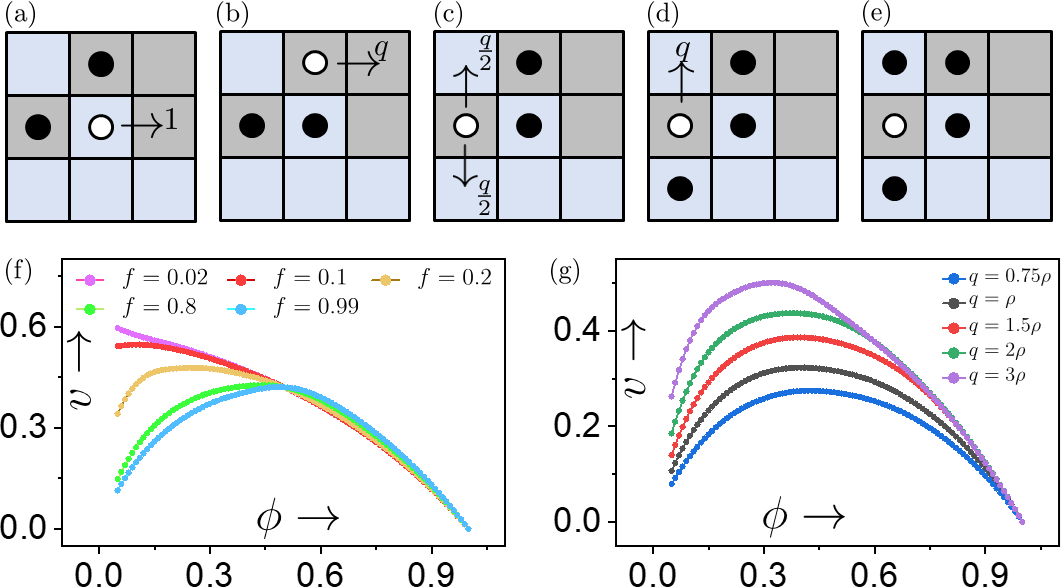}
        \caption{
            (a) $\ldots$ (e) shows the schematic for the simulation rules. Here the particle marked white is chosen to move, while black particles remain stationary. Slow sites are marked gray, while fast sites are marked in light blue. For this proposed theoretical model of hard-core discs hopping between the sites of a two-dimensional lattice, (f) shows the average particle velocity $\langle \omega \rangle$ vs. packing fraction $\phi$ for different values of the parameter $f$ characterizing the hop rate distribution, with hop rate parameter $q= 2\phi$, see text.
            (g) By changing the parameter $k$ in $q= k\phi$, we can change the position and the height of peaks in the $\langle \omega \rangle$ versus $\phi$ curve. }
        \label{fig:simdata}
    \end{figure}

    The lattice constituting our model consists of multiple lanes placed adjacent to one another, and with periodic boundary conditions only in the longitudinal direction. Thus, referring to Fig.\ref{fig:simdata}(a) -- (e), periodic boundary condition applies as one moves from left to right along the lattice, while no periodic condition applies as one wraps around the lattice from top to bottom. We specifically consider $N$ hard-core discs on a two-dimensional lattice of $L_1$ rows and $L_2$ columns, with filling fraction $\phi =N/L$ and $L=L_1 L_2$. The dynamics of our model involves stochastic hopping of discs to vacant nearest-neighbor sites, either in the same or in the adjacent lane, with the preferential motion of rods accounted for by considering the hopping to be biased along one of the two longitudinal directions. We will here consider the extreme case of biased hopping in which hopping is possible only to the right, see Fig.~\ref{fig:simdata}(a)--(e). The results are expected to remain qualitatively unaffected on considering the motion to be possible both to the right and to the left, with a bias, say, to the right. A schematic diagram showing the rules of the simulation is shown in Fig. \ref{fig:simdata} (a-e). The model has spatially-distributed disorder in that every site has an associated hopping rate $\alpha$ for a disc that occupies the site, which we consider to be a random variable sampled from a bimodal distribution: $P(\alpha = q) = f, P(\alpha = 1) = 1-f$, with $0<f<1$. The lattice is generated with a fixed probability $f$ of having slow sites , and $1-f$ for fast sites. Slow sites are marked in gray in Fig. \ref{fig:simdata} (a,e). Referring to Fig.\ref{fig:simdata}, panels (a) and (b), the particle marked white and occupying a site with associated hopping rate $1$, panel (a), or, $q$, panel (b), is shown to move to the right neighboring site. Note that the particle move is allowed in the two cases since the destination site was empty before the move. If however the destination site happens to be occupied, as is the case in panel (c), then the white particle has the possibility to change lanes. A situation may arise when the destination sites in the two adjacent lanes are empty, in which case, the particle moves to either of the two sites with equal probability $q/2$ (respectively, $1/2$) if the particle before the move was on a slow site (respectively, a fast site). The former case is shown in Fig.\ref{fig:simdata}(c). Another situation may arise when only one of the two possible destination sites in the adjacent lanes are empty; in this case, the particle moves to this empty site with probability $1$ or $q$, depending respectively on whether the particle before making the move was on a fast or a slow site, see Fig.\ref{fig:simdata}(d). Figure \ref{fig:simdata}(e) represents the case when all possible destination sites are occupied, and hence the white particle cannot make any move whatsoever.
    In our experiments, we have observed that a rod makes a move provided another rod in its vicinity collides with it and pushes it forward. The probability to find another rod in the neighborhood grows linearly with the filling fraction of rods. This aspect is taken care of in our stochastic model by stipulating that the parameter $q$ is a linear function of $\phi$. The stochastic model proposed here is a nontrivial generalization of the paradigmatic asymmetric simple exclusion process \cite{Golinelli_2006}. The lattice is periodic in x direction, and the average current $\langle \omega \rangle$ is measured is measured in this direction.

    % \begin{figure}[t]
    % \includegraphics[width=.95\linewidth]{pic 5_5.png}
    % \caption{ \textcolor{red}{The schematic for the simulation are shown on a 8$\times$3 lattice. The lattice is generated with a fixed probability $f$ of having slow sites, $1-f$ for fast sites. A particle at a fast site, marked green hops forward with a probability 1, while particles at slow sites, marked red hop forward with a rate $q$ iff the site in front is empty. If the site in front of a particle is occupied, it can hop to an empty site in any one of the adjacent rows with appropriate hopping probabilities. The lattice is periodic in one direction only, and this is the direction in which all particles hop, and the average current $\langle \omega \rangle$ is measured.}
    % }
    % \label{fig:SIM_rules}
    % \end{figure}
    We studied our proposed model by performing Monte Carlo simulations of the dynamics, whereby evolution in discrete time steps involves sampling a particle at random and attempting to move it to a nearest-neighbor vacant site according to the rules mentioned in the preceding paragraph. These simulations were carried out for $L=100, 200, 500$, and different values of $\phi$. There were no appreciable differences on changing lattice size. The system was allowed to settle into a stationary state, after which measurements of average velocity $\langle \omega \rangle$ of individual discs were made (because of periodic boundary conditions, the velocity that one measures is actually the angular velocity). Our obtained results for $\langle \omega \rangle$ versus $\phi$ show the same characteristic rise and fall as in our experimental data; The position and the height of the peak depends on both $f$ and $q$ (Fig. \ref{fig:simdata}, panels (f) and (g)). We have also performed alternative simulations where particles were demarcated with probability $f$ to be slow with slow speed $q$, obtaining similar numerical results. This simulation scheme mirrors the experimental feature of imperfections between particles.
    
    The different translational and rotational degrees of freedom of
    an Euclidean object may get coupled in an external field, resulting in a dynamics that violates time-reversal symmetry. Toys such as the rattle back or the tippe top are single-particle examples of this phenomenon. The present study extends in a non-trivial and hitherto unexplored manner this domain of work to a multi-particle setup of interacting achiral objects, wherein chirality that drives a current in the system is an emergent property.

 Shamik Gupta acknowledges support from the Science and Engineering Research Board (SERB), India under SERB-TARE scheme Grant No. TAR/2018/000023, SERB-MATRICS scheme Grant No. MTR/2019/000560, and SERB-CRG scheme Grant No. CRG/2020/000596. \cor{SG acknowledges  support of the Department of Atomic Energy, Government of India, under Project No. 12-R $\&$DTFR-5.10-0100.}
    
% \bibliographystyle{apsrev4-1}
    %\printbibliogaphy
%\bibliography{chiral}

    % \end{linenumbers}
    
\end{document}